\documentclass[prl,twocolumn,showpacs,floatfix,amsmath,amssymb,superscriptaddress]{revtex4-1}
\usepackage{amsfonts}
\usepackage{stmaryrd}
\usepackage{bbm}
\usepackage{mathrsfs}
\usepackage{tipa}
\usepackage{amssymb}
\usepackage{txfonts}
\usepackage{graphicx}
\usepackage{dcolumn}
\usepackage{epstopdf}
\usepackage[colorlinks,linkcolor=blue,urlcolor=blue,citecolor=blue]{hyperref}
\usepackage{multirow}
\usepackage{subfigure}
\usepackage{url}

\begin{document}
\newcommand*{\cm}{cm$^{-1}$\,}
\newcommand*{\Tc}{T$_c$\,}

\title{Revealing ultra-strong magnon-photon coupling in a polar antiferromagnet Fe$_{2}$Mo$_{3}$O$_{8}$ by time domain terahertz spectroscopy}
\author{L. Y. Shi}
\affiliation{International Center for Quantum Materials, School of Physics, Peking University, Beijing 100871, China}

\author{D. Wu}
\affiliation{International Center for Quantum Materials, School of Physics, Peking University, Beijing 100871, China}

\author{Z. X. Wang}
\affiliation{International Center for Quantum Materials, School of Physics, Peking University, Beijing 100871, China}

\author{T. Lin}
\affiliation{International Center for Quantum Materials, School of Physics, Peking University, Beijing 100871, China}

\author{C. M. Hu}
\affiliation{Department of Physics and Astronomy, University of Manitoba, Winnipeg, Manitoba R3T 2N2, Canada}

\author{N. L. Wang}
\email{nlwang@pku.edu.cn}
\affiliation{International Center for Quantum Materials, School of Physics, Peking University, Beijing 100871, China}
\affiliation{Collaborative Innovation Center of Quantum Matter, Beijing 100871, China}

\begin{abstract}

Strong coupling between magnon and electromagnetic wave can lead to the formation of a coupled spin-photon quasiparticle named as magnon-polariton. The phenomenon is well studied for ferromagnetic systems inside microwave cavities in recent years. However, formation of magnon-polariton is rarely seen for an antiferromagnet (AFM) because the strong coupling condition is not easily fulfilled. Here we present time-domain terahertz measurement on a multiferroic polar antiferromagnet Fe$_{2}$Mo$_{3}$O$_{8}$. We find clearly beating between two modes at frequencies above and below the electric-active magnon frequency below $T_N$, which we assign to the formation of AFM magnon-polariton. An ultra-strong spin-photon coupling effect is derived based on the energy level splitting. However, the AFM magnon-polariton is absent in the frequency domain measurement. Our work reveals that the coherent magnon formation driven by the ultrashort THz pulse provides a new way to detect polariton mode splitting.
\end{abstract}

\maketitle

The strong coupling between light or electromagnetic wave and collective excitation of matter can produce a hybrid state or quasiparticle with energy levels different from either the electromagnetic wave or the collective excitation of matter. The hybrid state, being referred to as polariton, can be used to transfer information between the electromagnetic wave and collective excitation of matter, thus opening up new prospects for quantum control and information technology. Among different light-matter coupling quasiparticles, the magnon polariton, \emph{i.e.} the hybrid light and magnon excitation in magnetically ordered material, has been anticipated and studied for a long time. However, due to the difficulty of achieving the strong coupling condition between spins of a magnet and electromagnetic wave, a clear hybridization effect could be hardly seen by traditional optical measurement. The magnon polariton was mainly inferred from the reflectance change within a very narrow frequency range in the reststrahl bands \cite{Mills1974,Jensen1997}. The eminent mode splitting arising from strong coupling effect was realized only in recent years in ferromagnet (FM) systems (\emph{e.g.} the traditional FM yttrium iron garnet (YIG)) specifically inside a microwave cavity resonator where the magnon resonance frequencies were tuned via external magnetic field to approach the cavity resonance frequency \cite{PhysRevLett.104.077202,PhysRevLett.111.127003,PhysRevLett.113.083603,PhysRevLett.113.156401,PhysRevLett.114.227201,PhysRevB.95.094416,PhysRevB.99.134445,Zhang2017,Harder2018}. Possible applications have been proposed for spintronics and information processing technologies based on the realization of cavity magnon polariton \cite{PhysRevLett.114.227201,Huebl2015}.

Antiferromagnetic materials are expected to be more promising for future spintronics application, for an antiferromagnet (AFM) has much faster spin dynamics than a FM. The magnon in an AFM is in the terahertz frequency range due to the strong exchange interaction between the spin sublattices, whereas the magnon in a FM material is in the gigahertz frequency range. However, due to the absence of a net magnetic moment, the coupling of AFM to the electromagnetic fields of light becomes weaker. Furthermore, the higher resonance frequency of AFM magnon makes both the stimuli and probe much more difficult by conventional electronics method. For those reasons, few work was reported on the magnon-photon coupling in AFM systems so far \cite{Biaek2020}.

With the advance of ultrafast laser technique, time domain terahertz (THz) spectroscopy has emerged as a powerful tool to detect the dynamics of AFM magnons. The THz waves generated from fetosecond laser are phase stable electromagnetic pulses with time duration of $\sim$ 1 ps. The THz electric field E(t) can be detected by electric-optical (EO) sampling on a nonlinear crystal in the time domain. By Fourier transformation of E(t), the spectrum in frequency domain can be obtained. Time domain measurement has certain advantages. A particular advantage is that the dynamical responses from different excitations may be separated in different time windows. In this letter, we present time-domain terahertz measurement on a multiferroic polar AFM Fe$_{2}$Mo$_{3}$O$_{8}$. We find clearly beating phenomenon between two modes at frequencies slightly above and below the electric-active magnon frequency below $T_N$, which demonstrates the formation of AFM magnon-polariton in the propagation of terahertz electromagnetic wave. The spin-photon coupling strength, defined as the ratio of the split energy gap over the mode resonance frequency, reaches 8$\%$, which is about one order of magnitude higher than the FM YIG inside a microwave cavity resonator\cite{PhysRevB.99.134445}. The ultra-strong coupling effect may be linked to the electric-active nature of this AFM magnon. We further show that the AFM magnon-polariton is absent in the frequency domain measurement by using Fourier transform infrared (FTIR) spectrometer. The study illustrates that the THz time domain technique provides a new way to detect polariton mode splitting.

\begin{figure}[htbp]
	\centering
	\includegraphics[width=8cm]{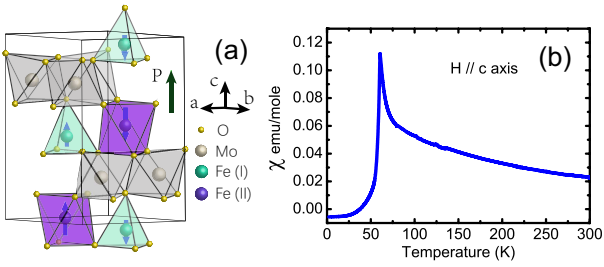}\\
	\caption{(a) Crystal and magnetic structure of Fe$_{2}$Mo$_{3}$O$_{8}$.  (b) The temperature dependent magnetization measurement for \textbf{H}//c-axis at 0.2 T. }\label{Fig:1}
\end{figure}

Fe$_{2}$Mo$_{3}$O$_{8}$ is a well-known polar AFM magnet with the electric polarization along the c-axis. The single crystals of Fe$_{2}$Mo$_{3}$O$_{8}$ were grown by chemical vapor transport method similar to the previous reports\cite{STROBEL1982242,STROBEL1983329}. The resulted hexagonal crystals were characterized by X-ray diffractions, Laue patterns and magnetization measurement. The lattice and magnetic structure of Fe$_{2}$Mo$_{3}$O$_{8}$ in the AFM ordered phase are shown in Fig.\ref{Fig:1} (a). There are two types of Fe ions locating in octahedron and tetrahedron of O with different magnetic moments. The octahedrons and tetrahedrons share their corners to form a honeycomb lattice in the ab-plane. Figure \ref{Fig:1} (b) shows the magnetic susceptibility as a function of temperature for \textbf{H}//c-axis measured in a Physical Property Measurement System under the magnetic field of 0.2 T. The data are similar to the reported results \cite{Wang2015}, indicating an AFM phase transition at $T_N$ = 60 K with ordered moments along c-axis. Due to the lack of inversion symmetry, an non-zero electric polarization is present in the compound. A considerable electromagnetic effect was reported in the multiferroic state below $T_N$ \cite{Wang2015, PhysRevX.5.031034}.

We performed time domain THz measurement on the crystals in a home built spectroscopy system with the terahertz beam normal to the ab-plane of the crystals. Details of THz system setup and measurements are
presented in the Supplemental Materials \cite{SI}. Figure \ref{Fig:2} (a) shows the waveforms of the transmitted THz electric field at several selected temperatures. Below $T_N$, we find long live oscillations after the main peak, which corresponds to the coherent magnon mode at frequency 1.25 THz. This collective excitation mode was studied in previous reports, however the time domain THz waveform within a sufficiently long time delay was not presented \cite{PhysRevB.95.020405, PhysRevLett.119.077206, PhysRevLett.120.037601}. The polarization dependent measurements indicated that it is an electric active magnon of the antiferromagnetic spin system\cite{PhysRevB.95.020405}. In general, the magnon excitation can be induced by an effective interaction through $dS_i/dt=[S_i, H_{eff}]$, where $H_{eff}$ reflects the interaction (e.g. Heisenberg interaction). Here, the magnetic excitation possesses in-plane (or perpendicular to the c axis) oscillation of electric polarization, which was proposed to be induced by the inverse Dzyaloshinskii-Moriya interaction and/or single-site anisotropy \cite{PhysRevB.95.020405}. Because the mode is electric active, i.e. driven by the electric field rather than the magnetic field of the THz electromagnetic wave, the magnon strength is considerably higher than the usual AFM magnon triggered by the magnetic dipole interaction, where coherent oscillations are usually observed after subtracting the waveform above the N\'{e}el temperature \cite{PhysRevB.98.094414}.

\begin{figure}[htbp]
	\centering
	\includegraphics[width=8.5cm]{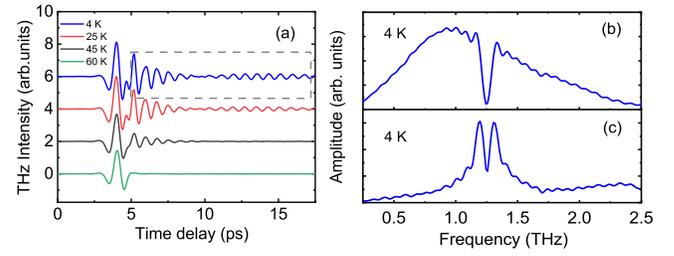}\\
	\caption{(a)  The time traces of the transmitted THz electric field at several selected temperatures.  The lines have been vertically shifted for clarity.  (b) Fourier transformation of the time trace of all the time range at 4 K, and (c) Fourier transformation of the time trace at 4 K after the main peak (in the time window of dashed rectangular from 5 ps to 17 ps).}\label{Fig:2}
\end{figure}

In our experiment, a clear beating phenomenon from two modes close in energy was observed in the time domain spectrum below $T_N$ as presented in Fig. \ref{Fig:2} (a). The signal intensity increases as the sample temperature decreases. We shall elaborate later that the beating can be explained to originate from the reversible energy exchange between THz wave and magnon mode, therefore it is an indication of the Rabi oscillation from hybridized magnon-polariton state. We can get the spectrum in frequency domain through Fourier transformation of E(t) to study the interaction between the magnon mode and THz electromagnetic wave through the beating frequencies. There are two ways to get the spectrum in the frequency domain. One is to perform Fourier transformation of the E(t) in the whole time delays. In this way we get the spectrum of a complete absorption, including the major absorption in the main peak and the absorption in the coherent collective excitation process after the main peak. The other one is to do the Fourier transformation of the time domain spectrum after the main peak. Then, the result is dominated by coherent collective modes. We perform Fourier transformation in both ways. The results for the lowest temperature at 4 K are shown in Fig. \ref{Fig:2} (b) and (c), respectively. In the frequency domain spectrum transformed from all time delays, we can see only one dip at frequency of 1.25 THz, related to the absorption of the magnon excitation. In the frequency spectrum transformed from the dashed rectangular in Fig. \ref{Fig:2} (a), which is the time window excluding the main peak, we can see two split peaks at 1.2 and 1.3 THz, respectively. Such beating phenomenon was rarely seen before. To the best of our knowledge, there was only one work reporting similar beating in TmFeO$_3$ \cite{Grishunin2018}. There were two related studies on the hybridized coupling between magnon mode of Fe$^{3+}$ and the electron paramagnetic resonance of Er$^{3+}$ in ErFeO$_3$ \cite{Li2018} or simultaneously hybridized coupling of magnon-phonon-polarition in specially designed LiNbO$_3$-ErFeO$_3$ chip devices \cite{Sivarajah2019}.

\begin{figure}[htbp]
	\centering
	\includegraphics[width=4cm]{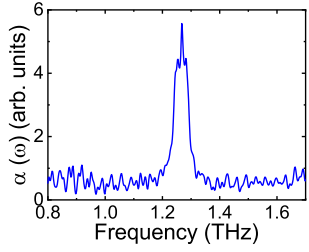}\\
	\caption{The high resolution absorption spectrum measured by Fourier transform infrared spectrometer. }\label{Fig:3}
\end{figure}

For a comparison, we also performed a frequency domain transmission spectroscopy measurement by using a FTIR spectrometer Bruker 80V for the sample at the lowest temperature 8 K. The absorption spectrum is shown in Fig. \ref{Fig:3}. We find a single narrow absorption peak at 1.25 THz. Within the signal-to-noise ratio and spectral resolution, no splitting is detected.

The different observations in the time domain THz and FTIR measurements are unexpected. It is of great interest to explore the underlying physics. As we mentioned in the introduction, the light matter interaction can lead to a hybrid state. Nevertheless, detecting the hybridized energy gap would depend on the coupling strength. Because the coupling between electromagnetic wave and magnetic excitation through magnetic dipole interaction is usually weak, it is not easy to observe hybrid excitation by traditional optical techniques. In recent years, much progress has been made for realizing the magnon polarition in FM systems inside microwave cavities. The cavities were used to block the escape of microwave and to limit the energy of microwave around the magnetic matter. In such cavity magnet system, the magnon excitation absorbs the energy from the microwave inside the cavity. The excited magnon will decay by re-emitting photons \cite{PhysRevLett.110.137204}. While the photons are limited in the cavity, they can be absorbed by the magnon again, and this leads to an energy circulation between magnon and microwave. In a real system, the energy may escape from the circulation. The imperfect mirrors will loss photons from the cavity, and the magnon may decay by other channels. If the energy exchange between magnon and microwave is more efficient than the escape of the microwave and the energy loss in the matter, the system is in a strong coupling regime, resulting in cavity magnon polariton.

\begin{figure}[htbp]
	\centering
	\includegraphics[width=8cm]{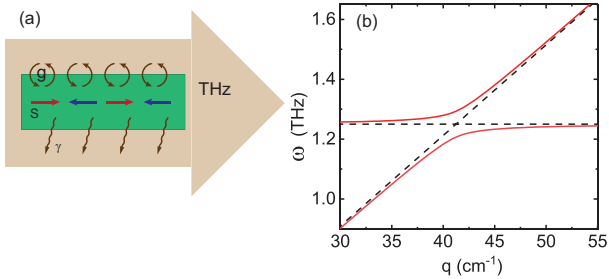}\\
	\caption{(a) A schematic diagram for the magnon-light interaction. s denotes the spin. A strong coupling is realized under the case when the energy transformation between magnon and THz photon (g) is more efficient than the escape of the THz wave and the magnon decay at a rate of $\gamma$. (b) A plot for the dispersion of the AFM magnon polariton from the real part of Eq. (2). The wave vector is defined as q=1/$\lambda$. The red curves indicate the coupled state. The two dashed lines indicate the dispersion of free photon and magnon. The two modes exhibit avoided crossing when approaching to the resonance frequency. }\label{Fig:4}
\end{figure}

For an AFM, the magnon appears at much higher energy, usually in THz frequency range. For this reason, the conventional cavity method based on the electronic technique in the microwave frequency regime is not applicable. Obviously, new techniques have to be developed to study the phenomenon. In the present time domain study, when the THz pulse generated by the ultrafast laser pulse transmits through the sample, as illustrated in Fig. \ref{Fig:4} (a), it transforms energy to the spin system and coherently excites the magnon. So the recorded main peak contains major absorption at the magnon frequency, the spectra in other frequencies would be transparent for the sample.  Because the processions of spins in the sample are triggered by the THz pulse, they have the same phase at the same wavefront of THz pulse, namely, their procession phase is locked to the phase of THz pulse. The oscillations after the main peak in Fig. \ref{Fig:2} (a) are the THz emissions from the coherent spin processions, i.e. magnon excitation in the time domain. The emitted THz wave frequency is equal to the magnon frequency, which can be absorbed again by the spin system to excite magnon. In this way the energy circling is built up, and a strong coupling condition is established. While the waves will eventually escape, there is no emitted THz signal at magnon frequency at sufficiently long time delay, then we can expect that overall absorption in the whole time delay comes from the magnon, as shown in Fig. \ref{Fig:2} (b). But in a proper time window of picoseconds time scale a beating emerges as a consequence of the strong coupling, reflecting two split modes with close frequencies, as in Fig. \ref{Fig:2} (c) . This is similar to a case of light interacting with a two-level system, resulting in a periodic change between absorption and emission of photons, which leads to Rabi splitting. Such coupled spin-photon system can be described by the two-level Hamiltonian \cite{PhysRevLett.114.227201,PhysRevB.99.134445,Sivarajah2019}.
\begin{equation}
H=H_p+H_m+H_{int}=\hbar
\begin{pmatrix}	
		q c & 0 \\
		0 & 0
	\end{pmatrix}
+\hbar
\begin{pmatrix}	
		0 & 0 \\
		0 & \omega_0-i\delta
	\end{pmatrix}
+\hbar
\begin{pmatrix}	
		0 & \Omega/2 \\
		\Omega/2 & 0
	\end{pmatrix}. 	\qquad
\label{chik}
\end{equation}
Here, $H_p$ and $H_m$ are the bare Hamiltonians of photon and magnon, $H_{int}$ denotes their coupling, $q$ the photon wavevector, c the speed of light in vacuum, $\omega_0$ the magnon frequency, $\delta$ the width of the magnon mode, $\Omega$ the splitting energy gap. The eigenfrequencies are
 \begin{equation}
\omega_{\pm}=\frac{q c +\omega_0}{2}-i\frac{\delta}{2}\pm\frac{1}{2}\sqrt{\Omega^2+((q c-\omega_0)+i\delta)^2} .
\label{chik}
\end{equation}
As a result, the coupling leads to the splitting of two branches between the magnon and the THz wave, i.e. magnon-polariton. The splitting frequency at the resonant frequency $\Omega$ is the Rabi frequency, being related to the coupling strength. A plot of the dispersion of the coupled magnon-polariton from Eq. (2) with $\omega_0$=1.25 THz is shown in Fig. \ref{Fig:4} (b). The two dashed lines in Fig. \ref{Fig:4} (b) indicate the dispersion of photon and magnon without coupling. The light-matter coupling leads to the avoided crossing at resonant frequency, as shown by the red line. The spin-photon coupling strength can be estimated from the ratio of the split energy gap over the mode resonance frequency, and it  reaches 8$\%$. This value is significantly higher than the value for FM YIG inside a microwave cavity resonator\cite{PhysRevB.99.134445}, reflecting an ultra-strong coupling effect.

\begin{figure}[htbp]
	\centering
	\includegraphics[width=8cm]{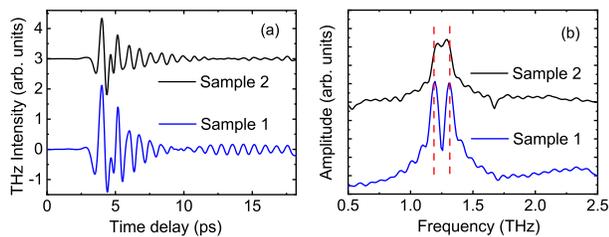}\\
	\caption{ (a) THz waveform for two samples with thickness of 1 mm (sample 1) and 0.4 mm (sample 2) in time domain. (b) The Fourier transformed spectra in the frequency domain for the two samples. The beating period and signal strength vary with the thickness of the samples. The dash lines indicate the splitting of the modes for the 1mm thickness sample. A smaller splitting is seen for the thinner sample.}\label{Fig:5}
\end{figure}

On the other hand, the splitting is not observed in FTIR experiment. In such experiment the continuous waves from the far-infrared light source, i.e. mercury lamp, transmit through the sample, and the far-infrared bolometer detector measures the power spectrum within a certain integration time. We believe that the coupling is still described by the same $H_{int}$, however, the light waves emitted by a mercury light source are incoherent in nature. Even though a similar beating should be formed in a time window of picoseconds for a specific array of wave passing through the sample, the average of different arrays of waves emitted by a mercury light source would give a zero amplitude in the oscillations. Additionally, the response and the integration time of bolometer detector is much longer than the time scale of tens of picoseconds, therefore, the Fourier transformation of the recorded far-infrared signal would give only a single absorption from the AFM magnon. This explains why the strong coupling polariton effect cannot be observed in FTIR measurement.

In fact, the coupling strength is enhanced for a assemble of N spins system. The coupling strength is known to be enhanced by a factor of $\sqrt{N}$ \cite{PhysRevLett.111.127003,PhysRevB.99.134445,H.Y.Yuan2017,Li2018}. In the present study, the number of spins interacting with the THz electromagnetic wave transmitting through the sample should depend on the thickness $l$ of the sample. Thus, we expect that the splitting of the modes $\Omega\sim\sqrt{l}$. To testify the relation between the number of spins and the coupling strength, we performed the experiments on samples with different thicknesses. Figure \ref{Fig:5} shows the THz spectra for two samples with thickness of 1 mm and 0.4 mm in both time and frequency domains at 4 K. Indeed, we observe a clear increase of beating period in the time domain and a reduction of the splitting in frequency domain for the thinner sample. The ratio of the splitting energy $\Omega$ of the two samples is about 0.6, being consistent with the ratio of the square root of thickness $\sim$0.63. The results confirm the positive correlation between spin number and coupling strength, one of the representative characteristics of magnon polariton coupling physics. We would also like to emphasize that the splitting does not depend on the pulse power of THz radiation. We performed power dependent measurement on a separate sample with different THz radiation powers in a THz time domain system with amplified laser system and confirmed that splitting does not change (the results are presented in the supplementary file \cite{SI}).

To summarize, we observed an ultra strong coupling effect between the THz electromagnetic wave and the electric active AFM magnon in Fe$_{2}$Mo$_{3}$O$_{8}$. The level splitting or Rabi oscillation is invisible in frequency domain measurement but can be monitored by time domain THz spectroscopy experiment in a proper time window. A clear beating between two modes at frequencies slightly above and below the electric-active magnon frequency is observed below $T_N$. We elaborate that the coherent formation of the spin procession enables the energy circulation in the propagation of the THz electromagnetic wave in the picoseconds time scale, thus establishing the strong coupling between the AFM magnon and THz light without a cavity. The coupling strength is related to the assemble of spins, therefore, the thickness of samples. Our study further illustrates that the electric active magnon in multiferroics with strong electromagnetic effect is a particularly suitable candidate to achieve the strong coupling of AFM magnon polarition, and the THz time domain technique offers new route to directly monitor the energy gap splitting.


This work was supported by National Natural Science Foundation of China (No. 11888101), the National Key Research and Development Program of China (No. 2017YFA0302904, 2016YFA0300902).

\bibliographystyle{apsrev4-1}
\bibliography{Fe2Mo3O8_mp}

\end{document}